\documentclass[sigconf]{acmart}
\AtBeginDocument{%
  }

\setcopyright{acmlicensed}
\copyrightyear{2026}
\acmYear{2026}
\acmDOI{XXXXXXX.XXXXXXX}
\acmConference[SIGSPATIAL '26]{The 34th ACM SIGSPATIAL International Conference on Advances in Geographic Information Systems}{November 3--6, 2026}{Riverside, CA, USA}
\acmBooktitle{The 34th ACM SIGSPATIAL International Conference on Advances in Geographic Information Systems (SIGSPATIAL '26), November 3--6, 2026, Riverside, CA, USA}

\begin{document}

\title[Resistance-Ranked versus Betweenness Pruning on Road Networks]{When Edge Importance Disagrees: Resistance-Ranked versus
Betweenness Pruning on Urban Road Networks}


\author{Jianru Shen}
\affiliation{%
  \institution{University of Montana}
  \city{Missoula}
  \state{Montana}
  \country{USA}}
\email{js258133@umconnect.umt.edu}


\begin{abstract}
Sparsifying large road-network graphs by removing a fraction of their edges
is a common way to reduce storage and accelerate spatial queries, but the
choice of which edges to remove can quietly degrade the network that
remains. We compare three edge-importance criteria for deciding what to
prune, namely uniform random removal, low edge-betweenness removal, and
low-effective-resistance removal, on 27 urban road networks drawn from
OpenStreetMap that span a wide range of sizes and street-network
morphologies. At edge-removal ratios from five to thirty percent we evaluate
each criterion along three complementary axes: length-weighted routing utility, a topology-only measure given by the retention of the second-smallest eigenvalue of the normalized Laplacian of the largest connected component, and the fraction of an externally defined arterial backbone, taken
from OpenStreetMap road classes, that each criterion destroys. The criteria
disagree. Low-effective-resistance ranking, which targets electrically redundant
edges, preserves connectivity well only below ten percent removal; beyond this level, retention falls below that of random removal in a
growing majority of cities, and by thirty percent the method removes a larger
share of the arterial backbone than random deletion does. Betweenness removal
is consistently the best of the three on all three axes, at every removal
ratio. Resistance-based redundancy is therefore not a safe proxy for routing
importance when compressing real road networks, and a routing-oriented
criterion is the more dependable choice.
\end{abstract}

\ccsdesc[500]{Information systems~Geographic information systems}
\ccsdesc[300]{Mathematics of computing~Graphs and surfaces}
\ccsdesc[300]{Mathematics of computing~Spectra of graphs}
\ccsdesc[100]{Theory of computation~Sparsification and spanners}


\maketitle

\section{Introduction}

Road networks are naturally represented as graphs whose nodes are
intersections and whose edges are road segments weighted by physical length.
As these graphs grow to metropolitan scale, many spatial systems reduce their
size by sparsification, removing a fraction of the edges to lower storage cost
and to accelerate downstream tasks such as routing, indexing, and network
analysis. Sparsification is attractive because a road network contains many
edges that seem redundant, but whether the reduced graph still behaves like
the original depends entirely on which edges are removed, a choice that has
received far less scrutiny than the compression itself.

That choice rests on a notion of edge importance, and different notions need
not agree. Ranking edges by effective resistance and deleting the lowest first
removes the edges whose endpoints are most strongly connected by alternative
routes; ranking by edge betweenness and deleting the lowest first instead
removes the connections that carry the least shortest-path traffic. The first
criterion is electrical, the second path-based, and on the nearly planar,
strongly arterial graphs that describe road networks it is not obvious whether
an edge judged redundant by one is also unimportant for the other. We test
this empirically by comparing three edge-removal criteria, uniform random
deletion, low-effective-resistance deletion, and low-betweenness deletion, on
27 urban road networks from OpenStreetMap spanning sizes from under one
thousand to nearly fifty thousand nodes and four street-network morphologies.
Each is sparsified at removal ratios from five to thirty percent and evaluated
not by whether it optimizes its own objective, but by whether the reduced
network remains useful.

We make three contributions. First, we evaluate these criteria along three
complementary axes, two of which do not reuse betweenness in their definitions:
length-weighted routing utility, the retention of normalized algebraic connectivity of
the largest connected component, and the fraction of an arterial backbone
defined externally from OpenStreetMap road classes. Second, we show that
effective-resistance scoring is not a safe proxy for routing importance on
road networks: beyond about ten percent removal it preserves neither routing
nor connectivity better than random deletion, and by thirty percent it removes
more of the arterial backbone than random deletion does, with the failure
visible on all three axes rather than only on a routing metric. Third, we show
that low-betweenness deletion is the dependable default for compressing a road
network while keeping it useful for routing, giving practitioners a concrete
and easily applied recommendation.

\section{Related Work}
Effective resistance, which views the graph as an electrical network, is a
classical edge-importance quantity~\cite{KleinRandic} that is small when an
edge's endpoints are joined by many parallel paths. It is central to spectral
sparsification: sampling edges according to their leverage scores $w_e R_e$ and reweighting the survivors yields a sparsifier that
provably approximates the Laplacian quadratic form, and hence the spectrum, of
the original graph~\cite{SpielmanSrivastava, BatsonSpielmanSrivastavaTeng}.
That guarantee is a property of the randomized, reweighted construction. The deterministic heuristic studied here ranks edges by effective resistance and deletes the least-resistant ones outright, without reweighting; this is not the sparsifier of \citet{SpielmanSrivastava} and inherits none of its guarantees. A systematic comparison by~\citet{HamannLindner} evaluates such methods on social networks. Work on the metric backbone of weighted networks shows that sparsifiers which do not preserve it alter shortest-path structure, and reports effective-resistance sampling among the global methods that do not preserve all shortest paths~\cite{SemiMetric}. These studies use social, biological, or synthetic
networks and evaluate properties such as diameter, community structure, or
spreading dynamics; none asks how common edge-removal criteria jointly affect
routing utility and structural preservation on real road networks. Edge betweenness, the fraction of shortest paths through an edge~\cite{Freeman, Brandes}, underlies routing-aware reduction; spatial work has reduced road-network size mainly to accelerate queries or for cartographic generalization~\cite{Geisberger}, not to compare how alternative criteria affect the network that remains.

\section{Experimental Setup}

\subsection{Road Networks}
We study 27 urban road networks in the United States, obtained from
OpenStreetMap with OSMnx~\cite{OSMnxBoeing}. For each city we download the
drivable street network and reduce it to a simple undirected graph whose
nodes are intersections and dead ends and whose edges carry the physical
length, in meters, of the corresponding road segment. Parallel and bidirectional edges between the same pair of nodes are collapsed to a single edge keeping the shortest length, with the merged edge taking the highest OSM road class among them, and we retain the largest connected
component so that every routing query is well defined on the original graph.
The networks span sizes from 838 nodes (Burlington, VT) to 48{,}612 nodes
(Phoenix, AZ), with a median of 11{,}462 nodes and a typical average degree
near 2.9, reflecting their nearly planar structure. To allow a check of
whether the findings depend on urban form, the cities are chosen to cover
four broad street-network morphologies: regular grids (for example Salt Lake
City, Denver, Indianapolis), historical cores (Boston, Savannah, Providence),
geographically constrained layouts (Seattle, Portland, Missoula), and
sprawling networks (Phoenix, Charlotte, Nashville).

Code and data to reproduce the main results are available at
\url{https://doi.org/10.5281/zenodo.20769521}.

\subsection{Edge-Removal Criteria}
Each criterion removes a target fraction of edges from a network, and we
evaluate the graph that remains. The three criteria differ only in the order
in which edges are removed. All edge scores are computed once on the original
graph, and the resulting fixed ranking is used at every removal ratio.

\emph{Random} removal deletes edges uniformly at random. Because a single
draw is noisy, we aggregate every reported quantity over ten random seeds,
combining per city before further analysis.

\emph{Betweenness} removal deletes edges in increasing order of edge
betweenness centrality~\cite{Freeman, Brandes}, removing the edges that lie on
the fewest shortest paths first. Betweenness is computed on the unweighted
topology, ignoring road length, so that any advantage of betweenness removal
cannot be a tautological consequence of optimizing the length-weighted routing
metric.

\emph{Effective-resistance} removal deletes edges in increasing order of
effective resistance, computed on the unweighted topology with unit
conductance on every road segment, removing the most electrically redundant
edges, those with many parallel alternatives, first. Computing exact effective resistance
requires the Laplacian pseudoinverse, which is infeasible at these graph
sizes, so we estimate all effective resistances with a Johnson--Lindenstrauss
projection~\cite{JohnsonLindenstrauss} using 50 random projection vectors,
which recovers each resistance from a small number of sparse Laplacian solves.
On a small network where exact values are available, the approximation has a median relative error of eight percent; we use the resulting estimates to rank edges.

\begin{figure*}[t]
  \centering
  \includegraphics[width=\textwidth]{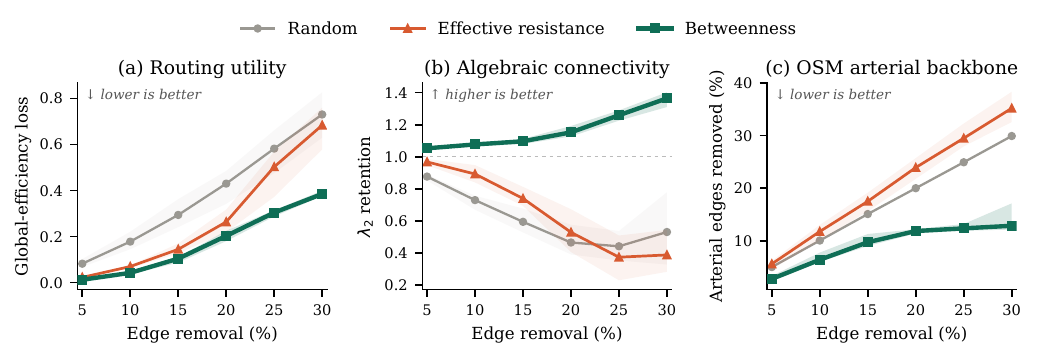}
  \caption{Preservation under increasing edge removal across 27 cities. Lines
show city medians, bands bootstrap 95\% confidence intervals; lower is better
in (a, c), higher in (b). Betweenness removal performs best on all three axes
throughout, while effective-resistance removal degrades beyond light
compression. The upturn in (b) near 30\% for random and effective-resistance
removal reflects fragmentation of the giant component, which raises the normalized $\lambda_2$ of the smaller surviving core.}
  \label{fig:three_axis}
\end{figure*}

\subsection{Removal Ratios and Evaluation}
Each criterion is evaluated at edge-removal ratios of 5, 10, 15, 20, 25, and
30 percent, computed as a fraction of the original edge count. This range
covers the light-removal regime, where resistance-based redundancy ranking should be most reliable, through the moderate compression relevant in practice. We
evaluate each sparsified network along three axes, chosen to provide complementary evidence without directly reusing the pruning scores.

\emph{Routing utility.} Shortest paths use physical road length as edge
weight, so routing reflects geographic distance. Computing all pairs is
infeasible on the larger networks, so we estimate routing quantities from a
fixed sample of 500 source nodes per city, reused across all removal ratios.
Writing $d_G(u,v)$ for the length-weighted shortest-path distance in graph
$G$ and $S$ for the sampled sources, the global efficiency~\cite{LatoraMarchiori}
\begin{equation}
E(G) = \frac{1}{|S|\,(n-1)} \sum_{u \in S} \sum_{v \neq u}
       \frac{1}{d_G(u,v)}
\end{equation}
averages the inverse shortest-path distance and stays well defined when some
pairs become disconnected, contributing zero. We report the relative
global-efficiency loss after sparsification, so that lower is better.

\emph{Normalized algebraic connectivity.} As a topology-only measure of how well
the global connectivity structure survives, we use normalized algebraic connectivity, defined as the second-smallest eigenvalue $\lambda_2$ of the unweighted normalized
Laplacian $\mathcal{L} = I - D^{-1/2} A D^{-1/2}$~\cite{Fiedler, ChungSpectral},
where $A$ is the adjacency matrix and $D$ the degree matrix. We compute
$\lambda_2$ with sparse Lanczos iteration on the largest connected component
of the sparsified graph, so that the measure reflects how well the surviving
backbone is connected rather than collapsing to zero the moment the graph
fragments; fragmentation is assessed separately through the disconnected-pair rate, the fraction of sampled source-target pairs left without a path, and is also reflected in the routing metric.
We report the retention $\lambda_2(G')/\lambda_2(G)$ relative to the original
graph, so that higher is better and values below one indicate lost
connectivity. Because this measure does not use edge lengths or shortest-path traffic, it provides a complementary structural view.

\emph{Arterial backbone.} As a third axis, defined without using the evaluated graph scores, we use the road hierarchy that OpenStreetMap assigns to each segment. We define
the arterial backbone of a city as the edges whose OSM class is motorway,
trunk, primary, or secondary, including their link ramps, and we measure the
fraction of this backbone that each criterion removes. Because the backbone is
defined externally by road class rather than by betweenness or any spectral
quantity, it gives an external check of whether a criterion preserves higher-order roads intended to support through movement; lower removal is better.

\section{Results}

Across all 27 cities the three criteria separate cleanly, and on two of the
three axes the separation is counterintuitive (Figure~\ref{fig:three_axis},
median across cities with a bootstrap 95\% confidence interval). We compare
criteria with two-sided Wilcoxon signed-rank tests across the 27 cities at
each ratio, treating each city as a single matched observation, which avoids
the inflated significance that arises when repeated measurements are pooled.

On routing utility (Figure~\ref{fig:three_axis}a), betweenness removal is best
at every ratio and by a widening margin: at 30 percent it loses a median of
0.385 of global efficiency, against 0.683 for resistance-ranked removal and 0.730
for random removal. Resistance-ranked removal holds a small advantage over random under light compression but converges to it, so beyond about 20 percent the method
built to remove redundant edges offers essentially no routing advantage over
random deletion. At 30 percent betweenness removal outperforms resistance-ranked removal in 26 of
27 cities and random removal in all 27 (both $p < 10^{-7}$).

On normalized algebraic connectivity (Figure~\ref{fig:three_axis}b), a topology-only measure complementary to routing, the behavior of resistance-ranked removal reverses
as compression grows. Under light removal it preserves connectivity best, as
its redundancy intuition predicts: at 5 percent its median $\lambda_2$ retention is 0.968, above random at 0.877. This does not survive. Beyond about
10 percent its retention falls steeply, dropping below random in a growing
number of cities, from 3 of 27 at 5 percent to 16 of 27 at 30 percent, where
its median retention is 0.388 against 0.531 for random. Betweenness removal, in
contrast, retains the normalized $\lambda_2$ of its largest component
above that of the original graph throughout, rising from 1.05 to 1.36 as it
sheds weakly connected peripheral edges; it exceeds resistance-ranked removal in 23
of 27 cities at 30 percent and random in 23 of 27 (both $p < 10^{-4}$). That resistance-ranked removal damages connectivity more than indiscriminate deletion, on a complementary topology-only measure, shows its weakness is structural rather than an artifact of how routing is scored. Betweenness removal also fragments the network least: at 30 percent removal its median disconnected-pair rate is 37 percent, against 63 percent under resistance-ranked removal and 75 percent for random removal.

\begin{figure*}[t]
  \centering
  \includegraphics[width=\textwidth]{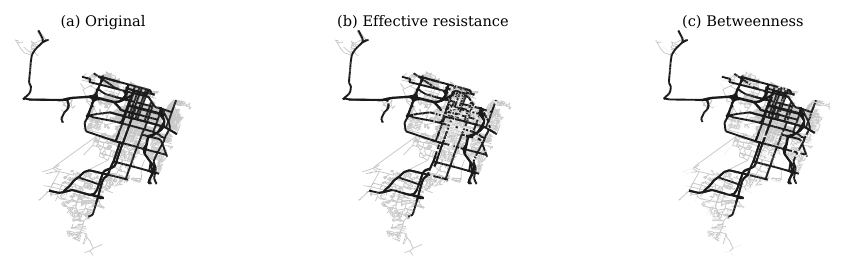}
  \caption{Savannah after 30\% edge removal. Effective-resistance removal
  deletes 32\% of the OSM arterial edges and disrupts the through-road
  skeleton; betweenness removal deletes 12\% and largely preserves it. Arterial
  roads dark, minor roads pale, removed edges nearly invisible.}
  \label{fig:savannah}
\end{figure*}

On the arterial backbone (Figure~\ref{fig:three_axis}c), defined entirely
outside the graph by road class, the pattern matches. At 30 percent removal,
resistance-ranked removal deletes a median of 35.2 percent of the backbone,
more than the 29.9 percent removed by random deletion, while betweenness
removal deletes only 12.8 percent. Betweenness removal preserves the backbone
better than both alternatives in all 27 cities (both $p < 10^{-7}$). These three axes converge on one explanation,
visible directly on the map (Figure~\ref{fig:savannah}). Our results suggest that many low-resistance edges lie in densely interconnected arterial regions rather than on peripheral streets. Effective-resistance ranking consequently deletes arterial edges disproportionately, which raises shortest-path distances, lowers algebraic connectivity, and destroys the road-class backbone; betweenness
removal does the opposite, shedding the peripheral edges that few shortest
paths use and leaving the arterial skeleton, and with it routing utility and
connectivity, intact.

\section{Discussion}

\paragraph{Practical recommendation.}
For practitioners the recommendation is direct. When the goal is to compress
an urban road network while keeping it useful for routing, low-betweenness
removal is the dependable default: across 27 cities it performs best on all three evaluated axes at every compression ratio we tested, leaving the arterial backbone almost entirely intact. Low-effective-resistance removal is not a reliable default for this purpose beyond light compression. Although its resistance-based redundancy score is appealing and does help under very light compression, beyond roughly ten percent removal it preserves neither routing nor connectivity any better than deleting edges at random, and it strips the arterial backbone at a higher rate than uniform random removal.

\paragraph{Interpretation.}
The broader lesson is that edge importance is not a single quantity. A score capturing one kind of importance, here the electrical redundancy of effective resistance, can mislead for a different purpose, here routing, when the two point in opposite directions on the structure that matters; in these networks, resistance-based redundancy often assigns low importance to edges with high routing value. An edge score should be validated against the task one actually cares about rather than trusted because it is principled in general.
\paragraph{Limitations.}
Several limitations bound these conclusions. First, we evaluate the
deterministic heuristic of deleting low-effective-resistance edges, which we study as a practical heuristic rather than the randomized, reweighted spectral sparsifier;
our results speak to that heuristic as an edge-importance criterion and not to
the guarantees of the sparsifier, and how a fully randomized construction
would behave on road networks is a separate question. Second, although routing utility is naturally more aligned with a shortest-path criterion, the other two axes do not reuse betweenness in their definitions; we rely on these topological and road-class axes to show that the effect is not an artifact of how routing is scored. Third, we study drivable road networks in the United
States, and whether the same decoupling holds for other spatial graphs, such
as pedestrian, transit, or utility networks, or in other regions, remains
open. Finally, the city-to-city variation is not explained by global graph
statistics; a structural account of when electrical redundancy and routing
importance diverge is the natural next step.

\bibliographystyle{ACM-Reference-Format}
\bibliography{references}


\end{document}